\documentclass{article}
\usepackage[a4paper, total={6in, 8in}]{geometry}
\usepackage{amsmath}
\usepackage{graphicx}
\usepackage{amssymb}

\usepackage{imakeidx}
\usepackage{framed}
\usepackage{pdfpages}
\usepackage{esint}
\usepackage{amsthm}
\usepackage{chemarr}
\usepackage{fancybox}
\usepackage{listings} 
\usepackage{multirow}
\usepackage{color}
\usepackage{lineno}
\usepackage{authblk}

\usepackage{empheq}

\newcommand{\f}{\mathbf{f}}
\newcommand{\g}{\mathbf{g}}
\newcommand{\Yv}{\mathbf{Y}}
\newcommand{\yv}{\mathbf{y}}
\newcommand{\xv}{\mathbf{x}}
\newcommand{\Xv}{\mathbf{X}}
\newcommand{\Pv}{\mathbf{P}}
\usepackage{gensymb}

\theoremstyle{definition}

\title{Stochasticity or Noise in Biochemical Reactions\footnote{This manuscript has been submitted for publication in the textbook {\em Quantitative Biology: Theory, Computational Methods and Examples of Models}, edited by Brian Munsky, Lev S. Tsimring, and William S. Hlavacek, to be published by MIT Press, \cite{Munsky:2017MIT}.}}
\author[1]{Zachary Fox}
\author[1,2,$\dagger$]{Brian Munsky}
\affil[1]{School of Biomedical Engineering, Colorado State University, Fort Collins, CO}
\affil[2]{Department of Chemical and Biological Engineering, Colorado State University, Fort Collins, CO}
\affil[$\dagger$]{To whom correspondence should be addressed: munsky@colostate.edu}
\date{}
\begin{document}
\maketitle
\begin{abstract}
Heterogeneity in gene expression across isogenic cell populations can give rise to phenotypic diversity, even when cells are in homogenous environments. This diversity arises from the discrete, stochastic nature of biochemical reactions, which naturally arise due to the very small numbers of genes, RNA, or protein molecules in single cells. Modern measurements of single biomolecules have created a vast wealth of information about the fluctuations of these molecules, but a quantitative understanding of these complex, stochastic systems requires precise computational tools. In this article, we present modern tools necessary to model variability in biological system and to compare model results to experimental data. We review the Chemical Master Equation and approaches to solve for probability distributions of discrete numbers of biomolecules. We discuss how to fit probability distributions to single-cell data using likelihood based approaches. Finally, we provide examples of fitting discrete stochastic models to single-molecule fluorescent in-situ hybridization data that quantifies RNA levels across populations of cells in bacteria and yeast.
\end{abstract}


\section{Introduction} \label{FoxIntro}
Our interest in the topic of stochastic biochemical reactions is driven by the fact that life is incredibly variable, even among genetically identical populations.  This diversity can arise from many sources, such as fluctuations in environmental stresses, nutrients, temperatures or other signals that affect cellular processes (e.g., growth, reproduction, death, etc.). But far more subtle inputs can also influence dynamics, and identical cells in controlled, seemingly homogenous environments can exhibit massive diversity.   In fact, diversity can arise from the rare and discrete nature of single-molecule events, especially as specific genes, mRNA molecules, or proteins interact with one another within the complex and unpredictable intracellular environment of chaotic fluctuations. Because a cell may have only one or two copies of a given gene, the unpredictable activation or deactivation of this gene can have radical effects.

\begin{figure}[h]
\begin{center}
\includegraphics[width=6in]{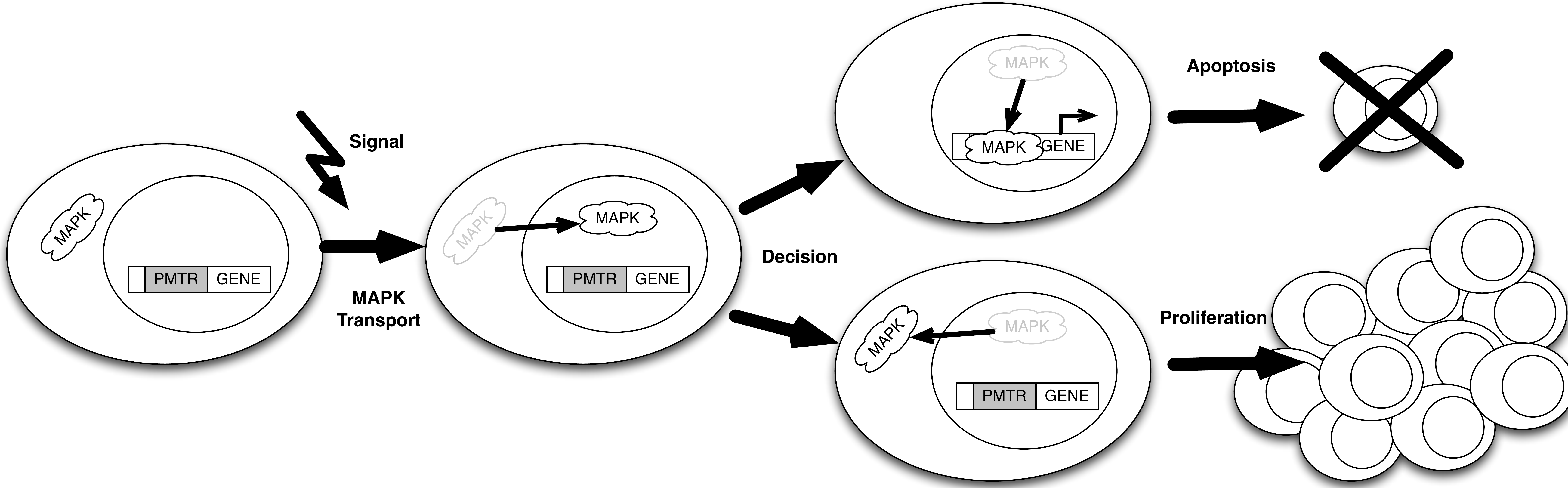}
  \end{center}
  \caption{\small {\bf Cartoon depiction of stochastic gene regulation.} (Left) The cell begins with a single gene and a single activated copy of a MAPK. (Top) Through a chance event, the MAPK molecule reaches and activates the gene, triggering events that lead to apoptosis. (Bottom) through a different chance event, the activator MAPK deactivates before it can activate the gene, and the cell is allowed to proliferate. \label{Munsky_Cartoon}}
\end{figure}

Figure \ref{Munsky_Cartoon} illustrates how randomness can cause phenotypical divergence of two cells with identical genetics and initial conditions.  To begin our exploration of such dynamics, we consider a simple model of a transcriptional response in exposure to a short period of external stress. This stress causes a temporary phosphorylation of a Mitogen Activated Protein Kinase (e.g., p38), which results in active transport of this kinase into the nucleus.  Once in the nucleus, two possibilities exist for this kinase's regulatory effect: (1) if the kinase reaches a specific gene promoter in time before it becomes de-phosphorylated, it could activate transcription of that stress response gene, or (2) if the kinase becomes de-phosphorylated and leaves the nucleus before it reaches the gene, that cell will remain inactive. In turn, the chance activation or inactivation of a given gene could have enormous effects on later processes \cite{McAdamsTrendsGenetics,Elowitz:2002,Thattai:2001,Hasty:2000ku,Ozbudak:2002,Fedoroff:2002,Kepler:2001}. For example, in processes related to oncogenesis, the MAPK signal may lead to apoptosis (programmed cell death) in one cell but allow proliferation of the other.  In general, the smaller the number of copies, the more variable will be the response, such that with one gene copy, switches may be all-or-nothing, but with more copies, the response can be much more graded. 

Different cellular mechanisms respond in different manners to cellular noise and fluctuations.  If variability is harmful to the organism, evolution will promote mechanisms, such as negative feedback \cite{Becskei:2000,Dublanche:2006,Nevozhay:2009,Huerta1998} or multistep proofreading processes \cite{hopfield1974kinetic,mckeithan1995kinetic,Bel:2010er,Swain:2002tc}, to decrease variability.  Conversely, discrete variations can also be used to cells' advantage.  For example, in combination with certain nonlinear processes, noise can amplify or dampen external signals \cite{Paulsson:2000} or help to improve the robustness of resonant behaviors \cite{Li:2005}.  The interplay of biochemical noise and non-linear dynamics can also result in stochastic switching, where genetically identical cells can express and occasionally change among multiple stable phenotypes \cite{Arkin1998,WolfArkin2002,Munsky:2005FOSBE,Tian:2006}. For single-cell organisms, this ability to switch at random can provide an evolutionary advantage to survive in uncertain environments \cite{Cagatay:2009}, but in other circumstances (such as cancer or autoimmune disease), the possibility to switch phenotypes can be deleterious (Fig. \ref{Munsky_Cartoon}).

Cellular variability can be measured using many experimental techniques, and there are many reviews in the literature discussing how single-cell and even single-molecule data can be collected through optical microscopy \cite{periasamy2013methods,yao2014chemistry,klein2014eight}, flow cytometry \cite{givan2013flow,aghaeepour2013critical}, and single-cell sequencing \cite{shapiro2013single,eberwine2014promise,grun2015design}.  One of the goals of quantitative biology is to use such data to infer predictive models for cellular behaviors. For example, the integration of stochastic models with experimental measurements makes it possible to understand of the dynamics of biochemical networks and to compare and contrast different possibilities for evolutionary design in the context of fluctuating environments \cite{Cagatay:2009}. Furthermore, cellular variability analyses can help to identify the mechanisms of cellular regulation \cite{Warmflash:2008,Dunlop:2008,Ronde:2009, Munsky:2012Science,Neuert:2013} and can help to identify quantitative, predictive models for cellular dynamics \cite{Munsky:2010IET,Munsky:2009MSB,Neuert:2013,Lou:2012NBT}.

But for these quantitative studies to succeed, we require computational tools that differ considerably from those utilized for deterministic analyses that utilize ordinary or partial differential equations.  These tools must take account of not just the temporal and spatial fluctuations, but also the random fluctuations (usually from unknown origins) that influence cellular dynamics. Some methods to allow for this treatment include kinetic Monte Carlo algorithms \cite{Gillespie:1977,gillespie2001,Allen:2005wy,Rao:2003,cao2005,Warmflash:2007} (discussed in Chapter 7 of \cite{Munsky:2017MIT}), linear noise approximations and moment closure techniques \cite{vanKampen,Elf:2003,Nasell:2003,GomezUribe:2007,Singh:2007BuB} (discussed in Chapter 6 of \cite{Munsky:2017MIT}), moment generating functions and spectral methods \cite{Sinitsyn2009,Walczak:2009} (discussed in Chapter 28 of \cite{Munsky:2017MIT}) and finite state projection approaches \cite{Munsky:2006JCP,Burrage:2006,Munsky:2008IEEE,Munsky:2008PHD} (discussed below in this article).  All of these stochastic computational analyses are derived from a common theoretical underpinning, which is known as the Chemical Master Equation or Forward Kolmogorov equation.\index{forward! Komogorov equation}  This article will focus on the derivation and explanation of the CME, and we will explore some simple CME analyses of stochastic gene regulation systems.

\section{The Chemical Master Equation}\label{FoxMesoscopic}\index{chemical master equation}
Biochemical reactions can be modeled at many different scales.  For example, in Chapters 3 and 4 of \cite{Munsky:2017MIT} explore macroscopic scales, where biochemical processes are treated with continuous-valued concentrations that evolve according to deterministic (ordinary, partial or algebraic) differential equations.  Alternatively, as in Chapter 10 of \cite{Munsky:2017MIT}, one could discuss molecular dynamics simulations, in which one could explore the motion and folding kinetics of individual biomolecules. Single-cell behaviors require an intermediate level of complexity, which we refer to as the {\em mesoscopic} scale.\index{mesoscopic scale}  For this scale, we describe each chemical species, $\{\mathcal{S}_1,\ldots,\mathcal{S}_N\}$, with non-negative integer vector, $\mathbf{x} = [\xi_1,\ldots,\xi_N]$, where $\xi_k$ is the number of molecules of the $k^{\rm th}$ molecular species. In this context, chemical reactions correspond to transitions from one state $\mathbf{x}_j$ to some other state $\mathbf{x}_i$. Such processes are typically represented using Markovian dynamics,\index{Markov! dynamics} which simply means that reaction rates are assumed to depend only upon the {\em current} state of the system -- {\em not} upon how the history or path by which that state has been reached.  This is known as the `well-mixed' assumption, and it can be justified from the analysis of bimolecular chemical kinetics in a well-mixed chemical solution (see \cite{gillespie1992} for a thorough derivation). However, the `well-mixed' result can also arise from the averaged influences of many un-modeled biochemical reactions, which may include complex biochemical reactions, such as transcription, translation, degradation, protein assembly and folding, all of which are comprised of numerous sub-steps (see \cite{Bel:2010er} for a discussion of this alternative origin of well-mixed kinetics).  This second, more permissive, origin of `well-mixedness' allow for reactions that are more complicated than bimolecular collisions and can result in more complex stochastic reaction rates, including Michaelis-Menten, Hill and other nonlinear functions.

For this article, we assume the most general Markov form for a discrete-value, continuous time chemical process.  
Each reaction in such a process is described by two quantities: the reaction's {\em stoichiometry vector} and its {\em propensity function}.\index{stoichiometry! vector}\index{propensity function}  The stoichiometry vector for the $\mu^{\rm th}$ reaction is represented by $\mathbf{\nu}_{\mu}$, and this vector describes how state changes as a result of the $\mu^{\rm th}$ reaction (e.g.,  $\mathbf{x} \rightarrow \mathbf{x}+\mathbf{\nu}_{\mu}$). For the reaction $s_1 \rightarrow s_2+s_3$, the stoichiometry vector is $\mathbf{\nu}=[-1,1,1]^T$.    The second part of a reaction's description is its {\em propensity function}, $w_{\mu}(\mathbf{x},t)dt$, which quantifies the probability the reaction will occur in a time step of length $dt$ given that the system starts in the state $\mathbf{x}$. Together, the stoichiometry and propensity functions for each reaction defines the stochastic dynamics. From these, one could simulate the stochastic process trajectories as discussed in Chapter 7 of \cite{Munsky:2017MIT}.  Instead, we use the stoichiometries and propensities to define a set of ordinary differential equations known as the chemical master equation (CME, \cite{McQuarrie}), which describes the probability distributions of the stochastic process.

To derive the CME, suppose that one knows the current probability mass for a particular state $\mathbf{x}_i$ at time $t$, which we write as, $p(\mathbf{x}_i,t)$.  Given our definitions of the stoichiometry and propensity functions, the probability that the system will be in the state $\mathbf{x}_i$ at time, $t+dt$, is equal to the sum of (\emph{i}) the probability that the system begins in the state $\textbf{x}_i$ at $t$ and remains there until $t+dt$, and (\emph{ii}) the probability that the system is in a different state at time $t$ and transitions to $\textbf{x}_i$ in the considered time step, $dt$.  This probability can be written as:
\begin{align}
\ p(\textbf{x}_i,t + dt)=p(\textbf{x}_i,t)\left(1 -
\sum_{\mu=1}^{M}w_{\mu}(\textbf{x}_i,t)dt \right) +
\sum_{\mu=1}^{M}p(\textbf{x}_i - \nu_{\mu},t)w_{\mu}(\textbf{x}_i -
\nu_{\mu},t)dt + \mathcal{O}(dt^2)\label{Foxincre_change},
\end{align}
where the first summation corresponds to the probability of starting and remaining in $\mathbf{x}_i$; the second summation corresponds to the probability of starting one reaction away from $\mathbf{x}_i$ (i.e., at $\mathbf{x}_i-\mathbf{\nu}_{\mu}$) and that particular reaction occurs, and the $\mathcal{O}(dt^2)$ term accounts for vanishingly small probability that more than one reaction occurs in the time step $(t,t+dt)$.  Subtracting $P(\mathbf{x}_i,t)$ from both sides and dividing through by $dt$ gives:
\begin{align}
\frac{p(\mathbf{x}_i,t + dt)-p(\mathbf{x}_i,t)}{dt}=
-\sum_{\mu=1}^{M}p(\mathbf{x}_i,t)w_{\mu}(\mathbf{x},t) +
\sum_{\mu=1}^{M}p(\mathbf{x}_i - \mathbf{\nu}_{\mu},t)w_{\mu}(\mathbf{x}_i -
\mathbf{\nu}_{\mu},t) + \mathcal{O}(dt)\label{Foxincre_change_2}.
\end{align}
Taking the limit as $dt\rightarrow 0$ yields the ordinary differential equation known as the CME:
\begin{align}
\frac{d}{dt} p(\mathbf{x}_i,t)=
-\sum_{\mu=1}^{M}p(\mathbf{x}_i,t)w_{\mu}(\mathbf{x},t)  +
\sum_{\mu=1}^{M}p(\mathbf{x}_i - \mathbf{\nu}_{\mu},t)w_{\mu}(\mathbf{x}_i -
\mathbf{\nu}_{\mu},t) \label{FoxCME_derived}.
\end{align}
We illustrate this equation in the following simple example of unregulated gene transcription.

{\bf Example 1}. {\em The CME for gene transcription}.  One of the simplest Markov systems is the birth/death process\footnote{This simple model provides an accurate description of transcription dynamics for many housekeeping genes (see review in \cite{Munsky:2012Science})}. This model has a single species, $\xi = [mRNA]$, and there are two reactions: (1) mRNA transcription has a stoichiometry of $\nu_1 = 1$ and a propensity function of $w_{1} = k$; and (2) degradation has a stoichiometry of $\nu_2 = -1$ and a propensity function of $w_{2} = \gamma \cdot \xi$.  If we let $i$ denote the number of mRNA, the CME for this process can be written 
\begin{align}
\frac{d}{dt} p_i(t)=
-p_i(t)(k+\gamma i) + p_{i-1}(t)(k) + p_{i+1}(t)\gamma(i+1).
\label{FoxCME_BD}
\end{align}

The CME is what is known as a linear ODE, but unlike the ODEs discussed in Chapter 3 of \cite{Munsky:2017MIT}, the dimension of the CME may be extremely large or even infinite, which means that it is usually impossible to solve exactly. For this reason simulation strategies or approximations are needed to analyze the CME, and several of these are discussed throughout various chapters of \cite{Munsky:2017MIT}. For a few rare models, analytical solutions are possible, and it is instructive to consider one such example:

{\bf Example 2}. {\em The steady state mRNA distribution for housekeeping genes.} Consider the CME given in the previous example (Eqn.\ \ref{FoxCME_BD}). The steady state distribution for this process is given by setting the time derivative to zero, which yields the following expression:
\begin{align}
\left. \frac{d}{dt}p_i\right|_{\rm SS} = -p_{{\rm SS}_i}(k+\gamma i) + p_{{\rm SS}_{i-1}}(k) + p_{{\rm SS}_{i+1}}\gamma(i+1) = 0,
\label{FoxCME_BD_SS}
\end{align}
which we can rewrite as:
\begin{align}
p_{{\rm SS}_{i+1}} = \frac{(k+\gamma i)p_{{\rm SS}_i} -(k)p_{{\rm SS}_{i-1}}}{\gamma (i+1)}\text{, for }i=\{0,1,2,\ldots\}.
\end{align}
For $i=0$, this yields:
\begin{align}
p_{{\rm SS}_1} = \frac{k}{\gamma}p_{{\rm SS}_0};
\end{align}
for $i=1$, the expression becomes:
\begin{align}
p_{{\rm SS}_2} &= \frac{(k+\gamma)p_{{\rm SS}_1} -(k)p_{{\rm SS}_0}}{2\gamma} = 
 \frac{(k+\gamma)(k/\gamma)p_{\rm SS} -(k)p_{{\rm SS}_0}}{2\gamma}=\frac{k^2}{2\gamma^2}  p_{{\rm SS}_0};
\end{align}
and in general
\begin{align}
p_{{\rm SS}_i} &=\frac{k^i}{i!\gamma^i}  p_{{\rm SS}_0} = \frac{\alpha^i}{i!}p_{{\rm SS}_0},
\end{align}
where $\alpha$ is the ratio of the transcription rate ($k$) and the degradation rate ($\gamma$). Because $p_{{\rm SS}_i}$ describes a probability distribution for the number of mRNA per cell, the sum of this series over all possible values of $i$ must be equal to one. Therefore,
\begin{align}
\sum_{i=0}^{\infty} p_{{\rm SS}_i} &=\sum_{i=0}^{\infty}\frac{\alpha^i}{i!}  p_{{\rm SS}_0} = 1.
\end{align}
Using the identity $\exp(\alpha) = \sum_{i=0}^{\infty}(\alpha^i/i!) $, we can solve for $p_{{\rm SS}_0}$ as:
\begin{align}
 p_{{\rm SS}_0} = \frac{1}{\exp(\alpha)} = \exp(-\alpha),
\end{align}
and therefore steady state probability distribution is given by the distribution:
\begin{align}
p_{{\rm SS}_i} = \frac{\alpha^i}{i!}\exp(-\alpha).
\end{align}
This distribution, which is known as the Poisson distribution, is characterized with a mean and variance both equal to $\alpha$ (see Exercise \ref{exercises}.1).

Unfortunately, analytical solutions like this are only known for a very small number of stochastic biochemical reaction models, and in most cases the solution of the CME requires partial or approximate solutions.  For this reason, most stochastic analyses of biochemical reactions makes use of stochastic simulations as discussed in Chapter 7 of \cite{Munsky:2017MIT}. Other approaches have been developed to analyze certain important summary statistics (i.e., means, variances and other statistical moments) as described in Chapters 6 and 28 of \cite{Munsky:2017MIT}. In the next section, we discuss a more direct solution to the CME, known as the Finite State Projection approach.\footnote{We note that the choice of which method (or even which version of a given method) to use for a given stochastic model is not always clear. Sometimes stochastic simulations are the only possible way to proceed, whereas in other cases moments or FSP analyses may be much more straightforward.  The choice depends strongly upon the model and the goal of the modeling analysis. The reader is encouraged to familiarize themselves with many different methods and to try them out on different toy problems. This experience will be instrumental to help one build the intuition to choose which approach for a given biological system.}

\section{Analyzing Population Statistics with FSP Approaches.}\label{FoxFSP}
As discussed above, there are a number of experimental techniques to measure and quantify cell-to-cell variability.  In particular, many of these approaches such as flow cytometry and single-molecule fluorescence in situ hybridization (smFISH, \cite{Femino:1998ek,Raj:2008NM}) are capable of taking only images or measurements from a given cell at a single time point in its development.  With these particular approaches, it is not possible to measure temporal trajectories of individual cells. Rather, one measures many cells at many different times or conditions and uses these to establish histograms for the cells' population statistics at many points in time. To capture such data using a stochastic model, it is necessary to solve the CME at corresponding times and experimental conditions. 
In this section, we discuss one such approach to accomplish this task, namely the Finite State Projection (FSP) approach \cite{Munsky:2006JCP}.

\subsection{Notation for the FSP}
The description of the Finite State Projection approach requires some additional notation, which we adopt from \cite{munsky20121d}.  The state of the system at any time is described by the integer population vector, $\{\xi_1,\ldots,\xi_N \}\in \mathbb{Z}^N_{\geq 0}$, where $\mathbb{Z}^N_{\geq 0}$ represents the set of all positive integer vectors of dimension $N$. 
Each possible state can be assigned a unique index $i$, meaning that state $\mathbf{x}_i$ refers to the vector, $\mathbf{x}_i = [\xi_1^{(i)},\ldots,\xi_N^{(i)}]$. If we use this enumeration to define the probability mass vector $\mathbf{P}\equiv [p(\mathbf{x}_1),p(\mathbf{x}_2),\ldots ]$, we can write the CME in a matrix form:
\begin{equation}
\frac{d}{dt} \mathbf{P}(t) = \mathbf{A}\mathbf{P}(t).\label{Fox:MatrixCME}
\end{equation}
Here, the matrix $\mathbf{A}$ is known as the {\em infinitesimal generator matrix}\index{infinitesimal generator! matrix}, and its (mostly-zero) elements can be written as:
\begin{equation}
A_{ij}  = \left\{\begin{array}{c|l} -\sum_{\mu=1}^M w_{\mu}(\mathbf{x}_j)  & \text{for }i=j,\\
   w_{\mu}(\mathbf{x}_j) & \text{for (i,j) such that } \mathbf{x}_i=\mathbf{x}_j+\mathbf{\nu}_{\mu},\\  
   0 & \text{otherwise}
\end{array}\right\}.\label{Fox_Inf_Gen}
\end{equation}
We note that the dimension of $\mathbf{A}$ is the same as the number of possible $\mathbf{x}_i$, which is infinite for many biological models.  The FSP approach \cite{Munsky:2006JCP} enables one to truncate the CME into a finite dimensional linear ordinary differential equation with precisely known error bounds.

Let ${\mathbf{J}}=\{ j_1,j_2,\hdots,j_{N_{\rm FSP}}\}$ denote a finite set of indices, such that ${\mathbf{X}}_{\mathbf{J}}$ denotes the finite set of states $\{{\mathbf{x}}_{j_1},{\mathbf{x}}_{j_2},\ldots,{\mathbf{x}}_{j_{N_{\rm FSP}}}\}$. Let ${\mathbf{J}}'$ denote the complement of the set ${\mathbf{J}}$ (i.e., all of the other indices that were not included in $\mathbf{J}$). Furthermore, for any vector $\mathbf{v}$ with the dimension of the set $\mathbf{X}$, let $\mathbf{v}_{\mathbf{J}}$ denote the subvector of $\mathbf{v}$ whose elements are chosen and ordered according to $\mathbf{J}$, and let $\mathbf{A}_{\mathbf{IJ}}$ denote the submatrix of $\mathbf{A}$ such that the rows have been chosen according to ${\mathbf{I}}$ and the columns have been chosen according to ${\mathbf{J}}$. 
For example, if ${\mathbf{I}}$ and ${\mathbf{J}}$ are defined as $\{1,2\}$ and $\{3,1\}$, respectively, then:
\begin{equation}
\left[ \begin{array}{ccc} a&b&c\\d&e&f\\g&h&k
\end{array} \right]_{{\mathbf{IJ}}}=\left[ \begin{array}{cc} c&a\\f&d
\end{array} \right]. \nonumber \end{equation}
For convenience, we will let $\mathbf{A}_{{\mathbf{J}}}\equiv \mathbf{A}_{{\mathbf{JJ}}}$.  
With this notation, we are now ready to state the main result of the Finite State Projection approach \cite{Munsky:2006JCP,Munsky:2008IEEE}, which we present in the format as it was described in \cite{Munsky:2008IET,munsky20121d}. 

\begin{figure}[h]
\begin{center}
\includegraphics[width=6in]{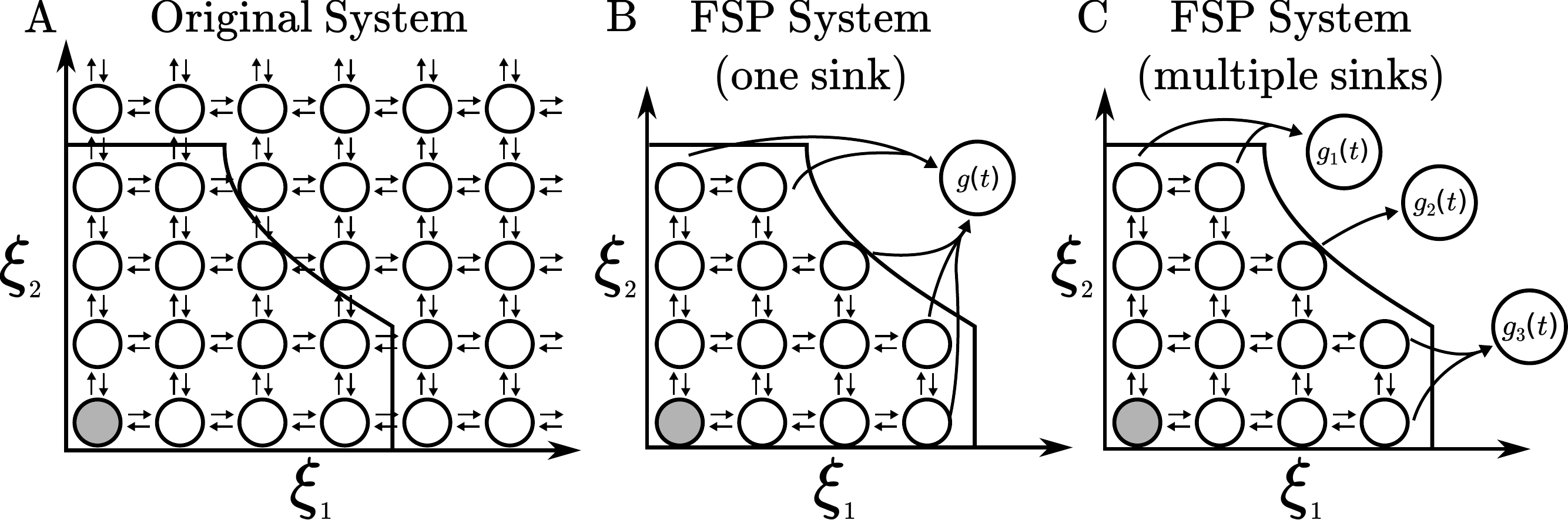}
\caption{\small A) {\bf A Markov chain depiction for a two-species reaction system.} Given an initial condition (indicated in grey), chemical reactions may increase or decrease the population counts of each of the two species $\xi_1$ and $\xi_2$.    
(a): A Markov chain, $\mathcal{M}$ for a chemically reacting system of two species. The process begins in the configuration shaded in grey and undergoes four reactions to increase/decrease the two different species populations. The dimension of the Master equation is equal to the total number of configurations in $\mathcal{M}$, which is infinite and therefore not amenable to an exact solution. (b) In the FSP algorithm, we select a finite configuration subset, $\mathbf{X}_{\mathbf{J}}$, and we collapse all remaining configurations in absorbing point masses $g$. This reformulation results in a finite dimensional Markov process, $\mathcal{M}_{\mathbf{J}}$. (c) The use of multiple absorbing sites makes it possible to keep track of how the probability measure leaves the projection space as described in \cite{Munsky:2008IET}.}
\label{FoxFSP_sinks_fig}
\end{center}
\end{figure}

We define the infinite state Markov process, $\mathcal{M}$, as the random walk on the configuration set $\mathbf{X}$, as shown in Fig.\ \ref{FoxFSP_sinks_fig}a.  The full, original master equation for this process is $\frac{d}{dt}{\mathbf{P}}(t) = \mathbf{A}(t)\mathbf{P}(t)$, with an initial distribution denoted as $\mathbf{P}(0)$.  We can define a new Markov process $\mathcal{M}_{\mathbf{J}}$ such as that in Fig.\ \ref{FoxFSP_sinks_fig}b, comprised of the configurations indexed by ${\mathbf{J}}$ plus one or more absorbing states.  The master equation of $\mathcal{M}_{\mathbf{J}}$ is given by
\begin{equation}
\frac{d}{dt}\left[\begin{array}{c} {\mathbf{P}}_{\mathbf{J}}^{\rm FSP}(t) \\ {g}(t) \end{array}\right] = \left[\begin{array}{cc} \mathbf{A}_{\mathbf{J}} &\mathbf{0} \\ -\mathbf{1}^T\mathbf{A}_{\mathbf{J}} & 0\end{array}\right]\left[\begin{array}{c} {\mathbf{P}}_{\mathbf{J}}^{\rm FSP}(t) \\ {g}(t) \end{array}\right], \label{FoxMaster_Eqn}
\end{equation}
with initial distribution,
\[
\left[\begin{array}{c}{\mathbf{P}}_{\mathbf{J}}^{\rm FSP}(0) \\ {g}(0) \end{array}\right] = \left[\begin{array}{c}{\mathbf{P}}_{\mathbf{J}}(0) \\ 1- \sum \mathbf{P}_{J}(0) \end{array}\right].
\]

\subsection{Properties of The Finite State Projection}
The finite Markov chain $\mathcal{M}_{\mathbf{J}}$ is closely related to the original $\mathcal{M}$. First, the probability mass that is contained in $g(t)$ is the {\em exact probability} that the system has left the set $\mathbf{X}_{\mathbf{J}}$ at {\em any} time $\tau \in[0,t]$. Second, elements of the vector $\mathbf{P}_{\mathbf{J}}^{\rm FSP}(t)$ are the {\em exact joint probabilities} that the system ({\em i}) is in the corresponding states $\mathbf{X}_{\mathbf{J}}$ at time $t$, and ({\em ii}) the system has never left the set $\mathbf{X}_{\mathbf{J}}$ at any time $\tau \in [0,t]$. 

From these properties, we can extract several pieces of insight into the solution of the original CME solution by solving and analyzing the finite system for $\mathbf{P}_{\mathbf{J}}^{\rm FSP}(t)$. First, because $\mathbf{P}_{\mathbf{J}}^{\rm FSP}(t)$ is a more restrictive joint distribution than $\mathbf{P}_{\mathbf{J}}(t)$, it is guaranteed that $\mathbf{P}_{\mathbf{J}}(t)\geq\mathbf{P}_{\mathbf{J}}^{\rm FSP}(t)\geq \mathbf{0}$ for any $\mathbf{J}$.  Second, we can derive a precise quantification of the difference between $\mathbf{P}(t)$ and $\mathbf{P}^{\rm FSP}(t)$ as follows \cite{munsky20121d}:
\begin{align}
\left| \left[ \begin{array}{c} \mathbf{P}_{\mathbf{J}}(t) \\ \mathbf{P}_{\mathbf{J}'}(t) \end{array}\right] -\left[ \begin{array}{c} \mathbf{P}_{\mathbf{J}}^{\rm FSP}(t) \\ \mathbf{0} \end{array}\right]\right|_1
 &=\left| \mathbf{P}_{\mathbf{J}}(t) -\mathbf{P}_{\mathbf{J}}^{\rm FSP}(t) \right|_1 +\left| \mathbf{P}_{J'}(t)\right|_1,\nonumber \\
 &=\left| \mathbf{P}_{\mathbf{J}}(t)\right|_1 -\left|\mathbf{P}_{\mathbf{J}}^{\rm FSP}(t) \right|_1 +\left| \mathbf{P}_{J'}(t)\right|_1,\nonumber \\
 &=1 -\left|\mathbf{P}_{\mathbf{J}}^{\rm FSP}(t) \right|_1,\nonumber \\
 &=g(t).\label{FoxFSP_accuracy}
\end{align}
With this, we can compute the accuracy of the FSP solution compared to the original CME problem.  Third, the FSP solution monotonically approaches the CME solution as new indices are added to $\mathbf{J}$ (i.e., as more states are added to $\mathbf{X}_{\mathbf {J}}$) as was proven in \cite{Munsky:2006JCP}.

\subsection{The FSP Algorithm.}
The formulation above suggests an FSP algorithm \cite{Munsky:2006JCP}, which examines a sequence of finite projections of the CME.  For each projection set, one can obtain an accuracy guarantee using Eqn.\ (\ref{FoxFSP_accuracy}).  If this accuracy is insufficient, more configurations can be added to the projection set, thereby monotonically improving the accuracy.  The full algorithm can be stated as given in Box 1.

\begin{framed}
\noindent{\bf Box 1}:  {\bf The Finite State Projection Algorithm}

\begin{tabbing}
 \=\textbf{Inputs   } \=Propensity functions and stoichiometry
for all reactions.\\
\>\>Initial probability density vector, $\mathbf{P}(0)$.\\
\>\>Final time of interest, $t_f$.  \\
\>\>Total amount of acceptable error, $\varepsilon>0$.\\
\\
\>\textbf{Step 0} \>Choose an initial finite set of states, $\mathbf{X}_{J_o}$, for the FSP.\\
\>\>Initialize a counter, $i=0$.\\
\\
\>\textbf{Step 1} \>Use propensity functions and stoichiometry to form
$\mathbf{A}_{J_i}$. \\
\>\>Compute $g(t_f)$ by solving Eqn.\ \ref{FoxMaster_Eqn} \\
\\
\>\textbf{Step 2} \>If $g(t_f)\leq \varepsilon$, \textbf{Stop}.\\
\>\>$\mathbf{P}^{\rm FSP}(t)$
approximates $\mathbf{P}(t_f)$ to within a total error of
$\varepsilon$.\\
\\
\>\textbf{Step 3} \>Add more states to find $\mathbf{X}_{J_{i+1}}$.\\
\>\> Increment $i$ and return to \textbf{Step 1}.
\end{tabbing}
\end{framed}

Steps 0 and 3 in the FSP algorithm (Box 1) can be accomplished through several different means, as described in the following.  

\subsubsection{Initialization of the FSP}\label{sec:FoxInit}
There are several different approaches through which one may initialize the FSP projection space. In \cite{Munsky:2006JCP},  $\mathbf{X}_{J_0}$ was initialized to include only the single state corresponding to the initial condition of the stochastic process. For this specification, the initial projection space is usually too small to retain sufficient probability mass for much time.  In \cite{Munsky:2007mtsFSP}, it was proposed to initialize $\mathbf{X}_{\mathbf{J}_0}$ with a set of states determined through the generation of multiple stochastic simulations\footnote{See Chapter 7 of \cite{Munsky:2017MIT} for details on how to generate stochastic simulations.}.  More simulations lead to a larger and better initial guess for $\mathbf{X}_{\mathbf{J}_0}$, which later requires fewer iterations of the FSP algorithm before convergence to an acceptable error. Of course there is some tradeoff between the time spent generating stochastic trajectories and that spent checking and expanding successive FSP projections.

Here we review a third approach to both initialize and expand the FSP projection \cite{munsky20121d,Fox:2016bna}.  In this approach, the projection space is defined as the set of all positive integer vectors that satisfy a specific set of polynomial constraints:
\begin{equation}
\mathbf{X}_{\mathbf{J}} = \{\mathbf{x}_i\}\text{, such that }\{f_k(\mathbf{x_i})\leq b_k\}\text{ for all constraints }k=\{1,2,\ldots,K\}, \label{FoxBoundary}
\end{equation}
where the functions $\{f_k(\mathbf{x})\}$ are polynomials of the populations and where the constraints $\{b_k\}$ set the limits on those polynomial functions.  Together, each $\{f_{k},b_{k}\}$ define a surface in the N-dimensional lattice of possible states.  For example, in the analysis of a two species $\{\xi_1,\xi_2\}$ system, one could use the projection shape functions \cite{munsky20121d}:
\begin{align}
f_1 &= -\xi_1,\  f_2 = -\xi_2,\  f_3 = \xi_1,\ f_4 = \xi_2, \nonumber \\ 
 f_5&=\max(0,\xi_1-4)\max(0,\xi_2-4), \nonumber\\
 f_6&=\max(0,\xi_1-4)^2\max(0,\xi_2-4),\nonumber\\
   f_7&=\max(0,\xi_1-4)\max(0,\xi_2-4)^2. \label{FoxProjSize}
\end{align}
The first two constraints ($f_1$ and $f_2$) set the lower bound on the numbers of each species as $b_1$ and $b_2$, respectively. For example, the trivial specification of $b_1=b_2=0$ restricts the set $X_{\mathbf{J}}$ to be non-negative.  Similarly, the third and fourth constraints, $f_3$ and $f_4$ specify the max populations of each species, and the remaining constraints specify additional, more complex bounding surfaces to constrain the population numbers.   In practice, the specific forms of the various functions $f_k$ can be changed, but all constraints must be specified such that the set of states monotonically increases as a function of any $b_k$.  Once the functions $\{f_k\}$ have been specified, the next step is to run one or more stochastic simulations and record the set of all unique states that are attained in those simulations, $\mathbf{X}_{\rm SSA}$. The initial boundary values $\{b_k\}$ can then be specified as:
\begin{equation}
b_k = \max_{\mathbf{x} \in \mathbf{X}_{\rm SSA}} f_k(\mathbf{x}) \text{, for }k\in\{1,2,\ldots,K\}.
\end{equation}

Once the initial FSP projection set has been specified, it may be necessary to expand it as required for Step 3 in Box 1.  A naive approach to carry out this expansion would be to include all of the states that are reachable in one reaction from the current set \cite{Munsky:2006JCP}.  However, such an approach is usually highly inefficient because it often expands too quickly in some dimensions while too slowly in others. Instead, we can use the polynomial shape functions defined above to specify a more directed FSP expansion routine. For this, we design $K$ absorbing points $\{g_1,\ldots,g_K\}$ where each $g_{k}(t)$ collects the probability mass that exits $\mathbf{X}_{\mathbf{J}}$ through the surface defined by the polynomial constraint $f_k(\mathbf{x})=b_k$.  We then find all pairs of states $\mathbf{x}_{j_i}$ and $\mathbf{x}_{j'}=\mathbf{x}_{j_i}+\mathbf{\nu}_{\mu}$ such that $j_i \in \mathbf{J}$ and $j' \in \mathbf{J'}$.  In construction of the FSP Markov chain ($\mathcal{M}_{\mathbf{f},\mathbf{b}}$ in Fig.\ \ref{FoxFSP_sinks_fig}), we split the flow of probability into $\mathbf{x}_{j'}$ equally among all sinks $g_{l}$ such that $f_{l}(\mathbf{x}_{j'}) > b_{l}$. For convenience, we let $n_{j'}$ denote the number of constraints violated by state $\mathbf{x}_{j'}$.
Under this definition, the finite dimensional master equation for $\mathcal{M}_{\mathbf{f},\mathbf{b}}$ can be written as:
\begin{equation}
\frac{d}{dt}\left[\begin{array}{c} {\mathbf{P}}^{\rm FSP}_{\mathbf{J}}(t) \\ \mathbf{g}(t) \end{array}\right]
=\left[\begin{array}{cc} \mathbf{A}_{J}(t) & \mathbf{0} \\
 \mathbf{B} & 0 \end{array}\right]
 \left[\begin{array}{c} \mathbf{P}^{\rm FSP}_{\mathbf{J}}(t) \\ \mathbf{g}(t)\end{array}\right], \label{FoxFSP_1}
\end{equation}
where $\mathbf{A}_J(t)$ is constructed as in Eqn.\ \ref{Fox_Inf_Gen}, and $\mathbf{B}$ is constructed as:
\begin{equation}
B_{k,i}  =
 \sum \frac{w_{\mu}(\mathbf{x}_{j_i})}{n_{j'}} \text{, where the sum is over all } \{\mu,i\}\text{ such that }
 \mathbf{x}_{j'}=\mathbf{x}_{j_i}+\mathbf{\nu}_{\mu}\text{ and } f_k(\mathbf{x}_{j'})>b_k
\end{equation}

The integration of Eqn.\ \ref{FoxFSP_1} over time not only provides a bounded solution for $\mathbf{P}(t)$ with total error equal to the sum of $\mathbf{g}(t)$, but it also records the amount of probability mass that exited through each of the polynomial constraint surfaces to collect in $\mathbf{g}(t)$. This information can then be utilized to relax these constraints in Step 3 of Box 1. Specifically, for each $k^{\rm th}$ boundary constraint such that $g_k(t_f)>\varepsilon/K$, one can then systematically relax the corresponding constraints by increasing the value of $b_k$.

A slightly different approach to expanding the FSP was proposed in \cite{sidje2015solving}, which makes use of additional stochastic simulations starting at the current state space and smoothing and expanding in the direction of those simulations that leave $\mathbf{X}_{\mathbf{J}}$. Still other approaches to expand the FSP state space were recently reviewed in \cite{Dinh:2016}. The choice of which expansion routine work best and in which circumstances is problem dependent and is a topic of ongoing research.

\subsection{Further reductions to the Finite State Projection}

Several approaches have been formulated to gain dramatic improvements to the FSP efficiency with little or no reduction in its accuracy. One common approach is to split the time interval into multiple subintervals, so that one can consider smaller portions of the state space during each time increment and thereby reduce the total computational effort \cite{Munsky:2007mtsFSP,Burrage:2006,henzinger2009sliding}. 
Another common approach to reduce the computational effort of the FSP is to apply linear systems theory to approximate the probability mass vector dynamics, $\mathbf{P}(t)$, using a lower dimensional vector $\mathbf{q}(t)\equiv \mathbf{\Phi}\mathbf{P}(t)$. A couple examples of such reductions have been based on concepts of controllability or observability \cite{Munsky:2006CDC,Munsky:2008IET}, separation of time scales \cite{Slaven:2006JCP,Munsky:2007ACC}, Krylov subspace methods \cite{Burrage:2006} and methods closely related to finite element analyses in partial differential equations \cite{Munsky:2008IEEE,tapia2012adaptive}. A significant amount of recent work has also been devoted to reformulating the FSP analysis into quantized tensor train notation \cite{kazeev2014direct}, which provides several advantages especially regarding memory allocation. 

Each of these reduction to the FSP has led to large gains in improvement to the FSP efficiency.  In turn these advancements are allowing for a rapidly growing class of CME systems that can be solved directly and then compared with experimental data as will be discussed in the following section.

\section{Comparing CME models to single-cell data} 
In order to identify or validate the parameters and mechanisms of a CME model, such a model must be compared to single-cell data such as flow cytometry \cite{Munsky:2009MSB,Lou:2012NBT} or single-cell fluorescence microscopy.  For example, the technique of single-molecule mRNA FISH \cite{Femino:1998ek,Raj:2008NM} is often used to used to measure and count the numbers of single RNA molecules in individual cells, and this technique provides the perfect experimental data with which to constrain an FSP model \cite{Neuert:2013,Shepherd:2013AC,Munsky:2015Methods,Fox:2016bna}. Because the smFISH technique requires cell fixation, it can only measure a cell at a single combination of time and experimental condition, but it can allow for the measurement of thousands of cells at multiple times and multiple conditions \cite{Neuert:2013}. In this case, each single-cell measurement can be considered to be independent given the particular time point and conditions of the experiment, a fact which helps to alleviate some challenges in defining the likelihood of the experimental data given a specified CME model.  In the following discussion, we will use this fact to formulate the likelihood of data at a single time point and single condition, while keeping in mind that the total likelihood at all time points and all conditions is simply the product the likelihoods over the different time points and conditions.

Suppose that a set of single-cell measurements from a smFISH experiment consists of the discrete counts of mRNA molecules in each of $N_c$ cells, given by $\Yv = [\yv_1,\yv_2,\ldots,\yv_{N_c}]$. For a given model, the CME solution gives the probabilities of these observations as: $[P(\yv_1),\ldots,P(\yv_{N_c})]$.
Assuming each cell is independent, the likelihood of observing all measurements is simply a product over these probabilities,
\begin{align}
    \mathcal{L(\Yv)} &= \prod_{i=1}^{N_c} P(\yv_i) \\
    \log\mathcal{L(\Yv)}  &= \sum_{i=1}^{N_c} \log P(\yv_i). \label{Foxlike1}
\end{align}
However, to compare this data to FSP models more efficiently, we can collect and count how many times each $\yv_i$ is observed, which we denote as $z_i$. This is equivalent to binning the data vector $\Yv$ into unit bins corresponding to the discrete number of molecules measured, where the counts $z_i$ of each bin are the number of cells in state $\xv_i$. Only some of the states $\mathbf{x}_i\in \mathbf{X}$ will be observed, and we can define the indices of observed states as $\mathcal{I}_Z$ (i.e., $\mathbf{X}_{\mathcal{I}_Z}$ is the set of observed states.) By writing the data in this form, we can directly compute the likelihood of the single-cell data given the CME model (with parameters $\mathbf{\Lambda}$) as:
\begin{align}
    \log\mathcal{L(\Yv)}  &= \sum_{j\in {\mathcal{I}_z}}^{N_m} z_j \log P(\xv_j | \mathbf{\Lambda}). \label{Foxlhood}
\end{align}
Applying the fact that the FSP approximation gives a lower bound on the solution of the CME, we are assured that the FSP provides a lower bound on the likelihood of the data given the CME model:
\begin{align}
    \log\mathcal{L(\Yv)}  \ge \sum_{j\in {\mathcal{I}_z}}^{N_m} z_j \log  P_{\mathbf{J}}^{\rm FSP}(\xv_j | \mathbf{\Lambda}). \label{Foxlikelihood}
\end{align}
We note that if the set $\mathbf{X}_\mathbf{J}$ does not include every state from $\mathbf{X}_{\mathcal{I}_z}$, then the FSP-provided lower bound is trivial ($ \log\mathcal{L(\Yv)}  \ge -\infty$).   This fact, combined with measured experimental data, provides another approach to specify the initial projection set for the FSP algorithm (see Section \ref{sec:FoxInit}).

In addition to the lower bound in Eqn.\ \ref{Foxlikelihood}, an upper bound on the likelihood of the data given the model can also be computed by making use of the FSP bounds derived in Eqn.\ \ref{FoxFSP_accuracy} \cite{Fox:2016bna}:
\begin{align}
    \log\mathcal{L(\Yv)}  \le  
    \max_{\sum \varepsilon_i = \sum{g}\text{, }\varepsilon_i\ge 0}
    \left(\sum_{j\in {\mathcal{I}_z}}^{N_m} z_j \log  \left(P_{\mathbf{J}}^{\rm FSP}(\xv_j | \mathbf{\Lambda}) + \varepsilon_i \right)\right), \label{FoxlikelihoodUB}
\end{align}
where the maximization optimization can be found through application of a simple water-filling algorithm \cite{Fox:2016bna}. We note that in some cases, even when the lower bound on the likelihood is trivial, this upper bound be used during parameter searches to reject poor models with less computational effort \cite{Fox:2016bna}.

Another common metric with which to compare CME models to single-cell data is to use the Kullback-Leibler divergence (KLD) between the CME solution and the empirical data distribution $\lbrace z_i/\sum z_j \rbrace$:
\begin{align}
   KLD\left(\frac{z_i}{\sum z_j},P(\xv | \mathbf{\Lambda})\right)=
   \sum_{i \in \mathcal{I}_z} \frac{z_i}{\sum z_j} \left(\log \left(P(\xv_i | \mathbf{\Lambda})\right) - \log \left(\frac{z_i}{\sum z_j}\right)\right) \label{FoxKLD}
\end{align}   
In is easily shown that maximization of the likelihood of the data (Eqn.\ \ref{Foxlhood}) and the minimization of the KLD (Eqn.\ \ref{FoxKLD}) occur at the same parameter set $\mathbf{\Lambda}$. 

\section{Examples}
In this section, we explore several example studies of how to construct CME and FSP stochastic models and compare them to discrete single-cell data. 
We first look at an example of constructing a gene expression model with two species, $\{\xi_1,\xi_2\}$, that interact with nonlinear propensities such that the species show switching or toggling in some parameter regimes.\index{toggle switch}
We then consider two examples with experimentally measured smFISH data in bacteria \cite{Shepherd:2013AC} and yeast \cite{Neuert:2013}.
In each of these parameter estimation examples, stochastic dynamics for a single gene and RNA were described by models with multiple genetic states corresponding to different activation levels of the gene. Such genetic states may correspond to transcription factor binding, or chromatin modifications that affect the expression level of the RNAs. 

\subsection{Toggle Model} \label{FoxToggle}
\begin{figure}[h]
\begin{center}
\includegraphics[width=5.5in]{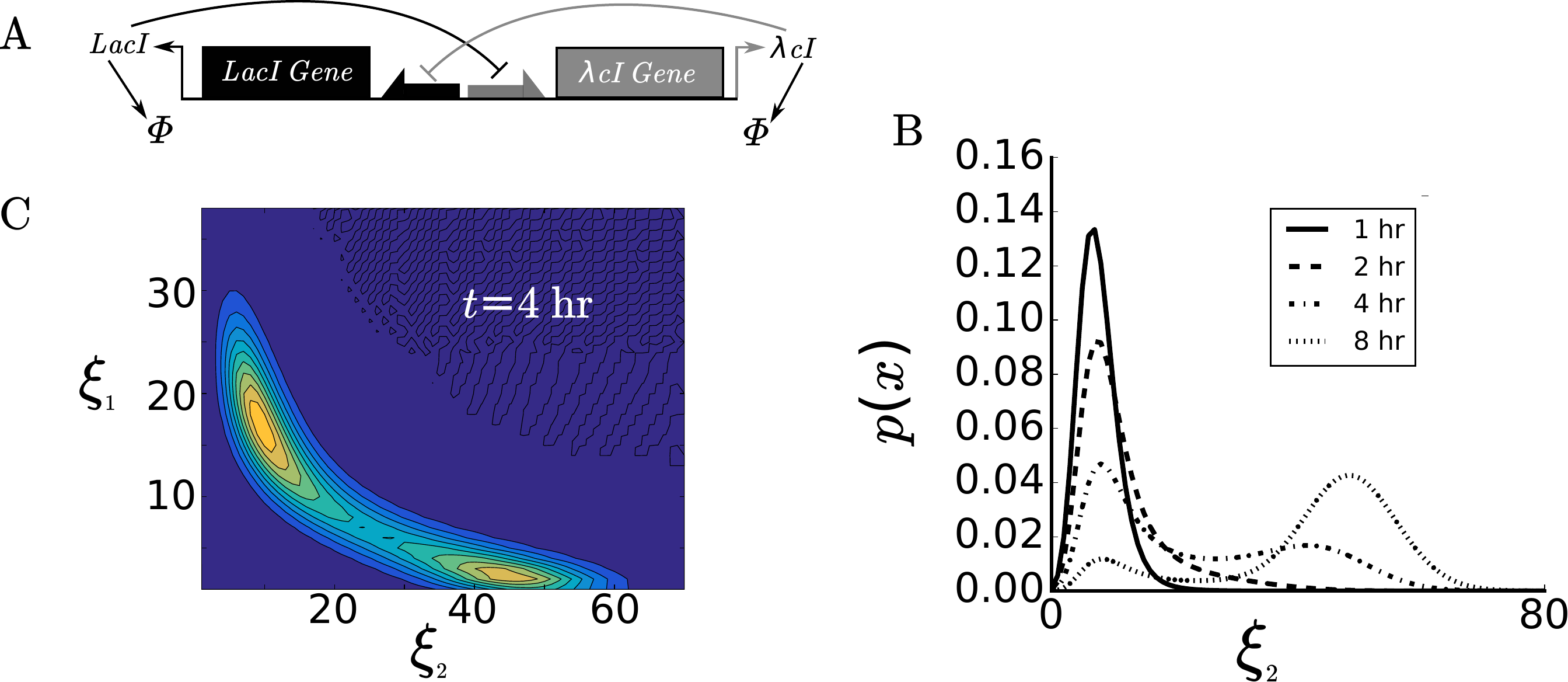}
  \end{center}
  \caption{\small (A) Demonstration of the two species toggle model. (B) Marginal distribution of LacI, (C) Joint distribution of the two species. In (B) and (C), parameters are given by $b_{\xi_2}=6.8e-5$, $b_{\xi_1}=2.2e-3$, $k_{\xi_2}=1.6e-2$, $k_{\xi_1}=1.7e-2$, $\alpha_{\xi_1}=6.1e-3$, $\eta_{\xi_1}=2.1$, $\eta_{\xi_2}=3.0$, $\gamma_{\xi_1}=\gamma_{\xi_2}=3.8e-4$. }\label{Foxtoggle_demo}
\end{figure}
Many different toggle switches have been constructed since Gardner et al \cite{Gardner:2000Nature}, and here we present a toggle model of two mutually inhibiting genes, \textit{$\lambda$cI} and \textit{lacI}, as shown in Fig.\ \ref{Foxtoggle_demo}A.
This example is a model similar to that provided by Tian and Burrage in \cite{Tian:2006}.
For this model, each state in the CME, given by $\xv_i = [\xi_1 \hspace{.2cm} \xi_2]_i$, corresponds to the discrete number of each protein, $\lambda$cI and LacI.
The toggle model therefore consists of four reactions, two ``birth" reactions, where each protein is created, and two ``death" reactions, where the proteins are degraded:
\begin{align*}
    \mathcal{R}1: \hspace{.2cm} \varnothing \xrightarrow{w_1} \xi_1; \hspace{.5cm} \mathcal{R}2: \hspace{.2cm} \xi_1 \xrightarrow{w_2}  \varnothing ;\\
    \mathcal{R}3: \hspace{.2cm} \varnothing \xrightarrow{w_3} \xi_2; \hspace{.5cm} \mathcal{R}4: \hspace{.2cm} \xi_2 \xrightarrow{w_4} \varnothing.
\end{align*}
The propensity functions $\mathbf{w} = \lbrace w_1,w_2,w_3,w_4 \rbrace$, are first-order for the degradation of each species, but the production of LacI is inhibited by $\lambda$cI (and vice versa). This inhibition is approximated by non-linear Hill equations,\index{Hill! function} 
\begin{align}
    w_1(\xi_2) &= b_{\xi_{1}} + \frac{k_{\xi_1}}{1+\alpha_{\xi_2}\xi_2^{\eta_{\xi_{2}}}}; \hspace{.5cm}
    w_2(\xi_1) = \gamma_{\xi_1} \cdot \xi_1; \nonumber \\
    w_3(\xi_1) &= b_{\xi_2} + \frac{k_{\xi_2}}{1+\alpha_{\xi_1} \xi_{1}^{\eta_{\xi_1}}}; \hspace{.5cm}
    w_4(\xi_2) = \gamma_{\xi_2} \cdot \xi_2.
\end{align}
Intuitively, if levels of LacI are high, then this will dominate the cell and inhibit $\lambda$cI production. The reverse case is also true - high levels of $\lambda$cI will be sustained via  repression of LacI expression.  Therefore, depending on the specific choice of parameters, either or both high expression states may be observed, as in Fig.\ \ref{Foxtoggle_demo}C. 
The matrix $\mathbf{A}$ for this system is written in terms of each propensity $w_i$ as:  
\begin{equation} \label{FoxAdef_toggle}
    \mathbf{A}_{ji}=\left\{ \begin{array}{c l}
    -\displaystyle\sum_{\mu=1}^4 w_{\mu}(\mathbf{x_i}) & \text{if $i=j$} \\
    w_1(\xv_i) & \text{for $i$ such that } \xv_j = \xv_i + [1,\hspace{.1cm} 0] \\
    w_2(\xv_i) & \text{for $i$ such that } \xv_j = \xv_i + [-1,\hspace{.1cm} 0] \\
    w_3(\xv_i) & \text{for $i$ such that } \xv_j = \xv_i + [0,\hspace{.1cm} 1] \\
    w_4(\xv_i) & \text{for $i$ such that } \xv_j = \xv_i + [0,\hspace{.1cm} -1] \\
    0 & \text{elsewhere}
    \end{array}  \right.
\end{equation}

To apply the FSP to the toggle model, we consider a subset of the constraint functions from Eqn.\ (\ref{FoxProjSize}), where $b_1$, $b_2 = N_{\xi_2}$ and $b_3 = N_{\xi_1}$ define the projection as
\begin{align*}
    \hspace{.2cm} \Xv_{\mathbf{J}} = \lbrace \xv_i \rbrace \text{ such that }  \left\{ \begin{array}{c l}
    f_1(\xv_i)=\max \{0,(\xi_2-4)(\xi_1-4)\}\leq b_1  \\
    f_2(\xv_i)=\xi_2 \leq b_2     \\
    f_3(\xv_i)=\xi_1 \leq b_3      
    \end{array} \right. \label{Foxshape}.
\end{align*}
By monotonically increasing $b_k$, more states are included, and the error $g(t)$ decreases to a specified tolerance $\varepsilon$.

The FSP approach to the toggle model may be used to calculate interesting dynamics to understand such biological switches. 
For example, in Exercise \ref{exercises}.8 we will strategically select placements of the sink state $g(t)$ to looking at the time it takes to switch between high levels of LacI or $\lambda$cI. 

\subsection{Matching stochastic models to small RNA measurements in bacteria}
Let us next consider a simple model of gene expression in \textit{Yersinia pseudotuberculosis}, an infectious bacteria species.
In \cite{Shepherd:2013AC}, the authors investigated the expression of a small (335 nt) RNA YSP8, which is expressed differentially at room temperature (\textasciitilde 25\degree C) compared to human body temperature (\textasciitilde  37\degree C), and therefore may play an important role in the bacteria's ability to infect humans.
Measurement from Shepherd et al \cite{Shepherd:2013AC} show characteristics of so-called bursting gene expression, in which many RNAs are made in a short period of time (with respect to the RNA degradation rate) and then the promoter turns to an OFF state in which no RNA are transcribed.
Thus, a natural model consists of two promoter states corresponding to active state and inactive states \cite{Munsky:2012Science,Munsky:2015PB}, where the {\em ysr8} promoter is either ON or OFF.  Transitions into an ON state occur with rate $k_{on}$ and turn off with rate $k_{off}$.  Transcription occurs with rate $k_{r}$.  Thus, the propensity vector is given by $\mathbf{w}=[k_{off},k_{on},k_{r}]$.  In this study, the authors found that the rate at which the gene activates is regulated by the temperature of the system, i.e.\ $k_{on}(T)$. 

\begin{figure}[h]
\begin{center}
\includegraphics[width=5in]{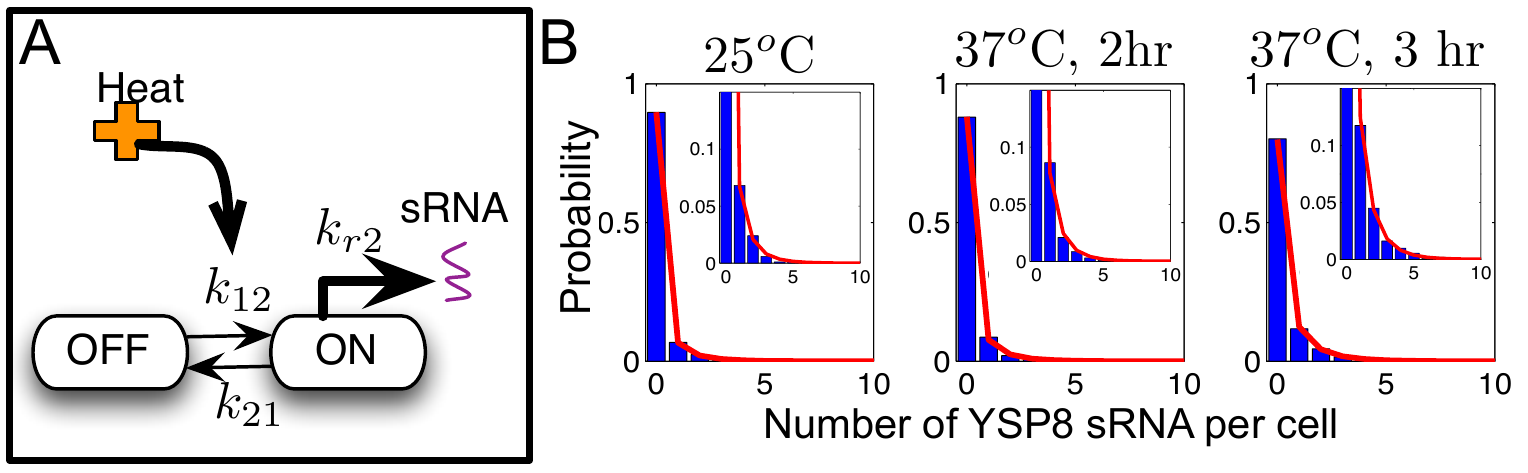}
\end{center}
\caption{\small Experimental and Computational Analyses of small RNA (YSP8) Transcription in {\it Yersinia Pestis} bacteria. Experimental data are shown in blue and model results are illustrated in red. A) Two-state model for the induction of YSP8 in response to temperature elevation from room temperature to human body temperature, which drives more cells into the activated `ON' state.  B) Distributions of smFISH data are shown by the bars and model fits are shown by the lines for a single time point before temperature change was applied, and then for two time points as the bacteria respond to the changing temperature. The insets highlight the changes in the tails (low-probability) events. Model parameters are given by $k_{r2}=k_{21}=1.19$, $\gamma = 1.00$, $k_{12}(25^\circ,t=2h)=0.138$, $k_{12}(25^\circ,t=2hr)=0.161$, $k_{12}(25^\circ,t=3hr)=0.286$.  Figure is adapted from \cite{Shepherd:2013AC} and reprinted from \cite{Munsky:2015Methods}, with permission from Elsevier.}\label{FoxYSP8}
\end{figure}

For this model, the infinitesimal generator can be broken down into three submatrices, corresponding to state transitions $\mathbf{T}$, transcription $\mathbf{B}$, and degradation $\boldsymbol{\Gamma}$:
\begin{equation}
    \mathbf{T(t)} = \begin{bmatrix}
                 -k_{on}(T) & k_{off} \\
                 k_{on}(T)   & -k_{off} 
                 \end{bmatrix}; \hspace{.5cm}
    \mathbf{B} = \begin{bmatrix}
                 k_r & 0 \\
                 0 & k_r \\
                 \end{bmatrix};\hspace{.5cm} 
    \boldsymbol{\Gamma} = \begin{bmatrix}  
                \gamma & 0 \\   
                0 & \gamma 
                \end{bmatrix}, \label{Foxtwo_state_subs}
\end{equation}
and the infinitesimal generator for the full CME $\mathbf{A}$ is given by a block tri-diagonal matrix.
\begin{align}
    \mathbf{A} = \begin{bmatrix}
                 \mathbf{T}-\mathbf{B} & \boldsymbol{\Gamma} &  \mathbf{0} & \ddots \\
                 \mathbf{B} &\mathbf{T}-\mathbf{B}- \boldsymbol{\Gamma} & 2 \boldsymbol{\Gamma}  &\ddots   \\
                 \mathbf{0} &\mathbf{B} &\mathbf{T}-\mathbf{B}- 2\boldsymbol{\Gamma}  &\ddots   \\
                 \vdots & \ddots & \ddots & \ddots
                 \end{bmatrix}
\end{align}
and the FSP version is 
\begin{align}
    \mathbf{A}_{\mathbf{J}}^{\rm FSP} = \begin{bmatrix}
                 \mathbf{T}-\mathbf{B} & \boldsymbol{\Gamma} &  \mathbf{0} & \hdots & \mathbf{0}\\
                 \mathbf{B} &\mathbf{T}-\mathbf{B}- \boldsymbol{\Gamma} & 2 \boldsymbol{\Gamma}  &\ddots & \mathbf{0}  \\
                 \mathbf{0} &\mathbf{B} &\mathbf{T}-\mathbf{B}- 2\boldsymbol{\Gamma}  &\ddots  & \mathbf{0} \\
                 \vdots & \ddots & \ddots & \ddots & \vdots \\ 
                 \mathbf{0} & \mathbf{0} & \mathbf{B} & \mathbf{T}-\mathbf{B}- N_m \boldsymbol{\Gamma} & \mathbf{0} \\
                 \mathbf{0} & \mathbf{0} & \mathbf{0} & \mathbf{1}^T\mathbf{B} & \mathbf{0} \\
                 \end{bmatrix}.
\end{align}

Figure \ref{FoxYSP8}B shows the resulting two-state  $k_{on}$-modulated model fit to the distributions of YSP8 in {\it Y. Pestis} at the initial steady state of 25$^o$C as well as at two and three hours post transition to 37$^{o}$C. Similar analysis was also applied to measurement of the YSR35 in {\it Y. pseudotuberculosis}, for which the same mechanism ($k_{on}$-modulated  regulation) also fit best to the measured distributions (see Reference \cite{Shepherd:2013AC}, Figure 5). Using the FSP algorithm, solving for the YSR8 transcript distributions takes an average of 0.0015 seconds to complete per parameter combination (on a 2.6 GHz Intel Core i7 Macbook Pro using {\sc Matlab}'s built in matrix exponentiation command ``{\it expm}'').

\subsection{Matching stochastic models stress response regulation in yeast}
\begin{figure}[th!]
\begin{center}
\includegraphics[width=5in]{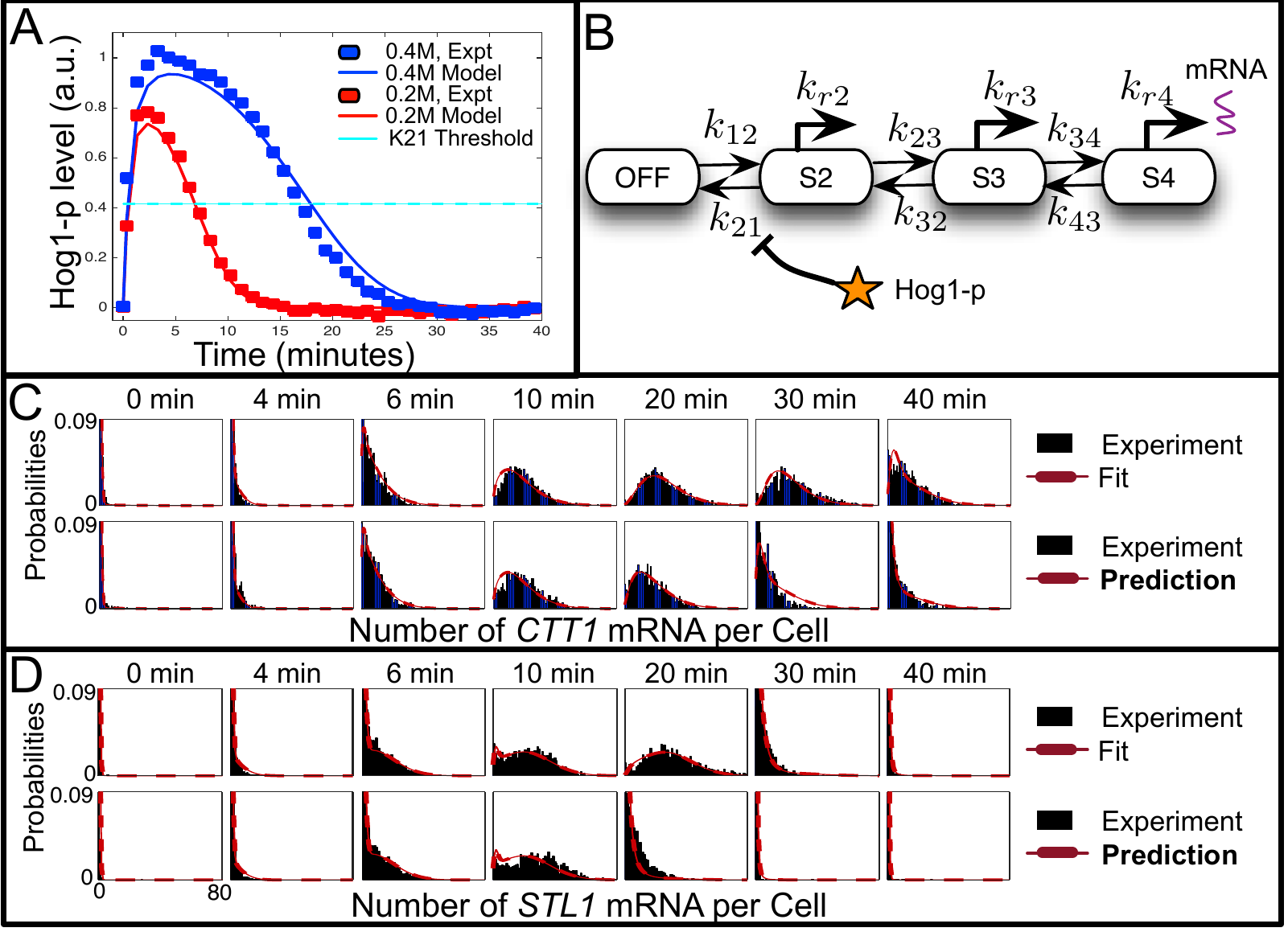}
\end{center}
\caption{\small  (A) Dynamics of Hog1-YFP during nuclear translocation in response to 0.2M NaCl and 0.4M NaCl osmotic stress measured with fluorescent time-lapse microscopy. If Hog1-p levels are past the horizontal line, the reaction $k_{21}$ does not occur, and the system is unable to return to an inactive state. (B) Identified four-state gene expression model, where Hog1-p in the nucleus inhibits the transitions from the active states (S2,S3,S4) to the inactive state (S1) for either the STL1 or CTT1 gene. (C,D) Measured smFISH data for CTT1 (C) and STL1 (D) are in the dark bar plots. The model fits that maximize the likelihood for the 0.4M NaCl gene expression data are on the top row for each gene in dashed lines and predictions for expression under 0.4M NaCl are on the bottom row for each gene. For model parameters, see Ref.\ \cite{Neuert:2013}. Figure is adapted from \cite{Neuert:2013} and reprinted from \cite{Munsky:2015Methods}, with permission from Elsevier. An example of fitting a stochastic model to similar data is explored in depth in Chapter 30 of \cite{Munsky:2017MIT}.}\label{FoxHogFig}
\end{figure}

In \cite{Neuert:2013}, full distributions of two genes activated in the high-osmolarity glycerol (HOG) MAPK pathway \cite{Brewster:2014}, STL1 and CTT1, were fit and predicted using the FSP analyses. 
Measurements of each gene were made using smFISH, and each shows transient expression as the yeast adapt to higher concentrations of salt in their environment. 
In such an environment, yeast activate the HOG-MAPK pathway, in which the Hog1-p kinase is phosphorylated, and translocates to the nucleus, where it activates a coordinated respone \cite{Muzzey:2009,Neuert:2013}. 
The dynamics of this pathway were measured at the single-cell level using a Hog1-p-yellow fluorscent protein fusion and fluorescence time lapse microscopy. 
The dynamics of the Hog1-p signaling are largely deterministic, as shown if Fig.\ \ref{FoxHogFig}A for two step inputs at 0.4M and 0.2M NaCl \cite{Neuert:2013}.
The authors designed smFISH probes to quantify expression of the different mRNA, and Figure \ref{FoxHogFig}C shows representative examples of the resulting distributions over time. 
These genes show significant variability from cell to cell. 

The authors then proposed a large class of models with different numbers of states of the gene and different mechanisms by which Hog1-p could affect transitions between gene states \cite{Neuert:2013}.
This class of models provided sufficient flexibility to match the data, where the more complex models (i.e.\ those that consisted of the most states) were able to fit the data quite well. 
The model parameters were identified by maximizing the likelihood of the smFISH data according to Eqn.\ (\ref{Foxlikelihood}).
Using a cross-validation approach, the authors identified the model that was most predictive, shown in Fig.\ \ref{FoxHogFig}(B). 
The optimal model consists of four gene states, where the stochastic transitions from the second state to the first state ({\it i.e.,} reaction rate $k_{21}$) is repressed by the time-varying Hog1-p signal.

This model is described similarly to the model in Eqn.\ (\ref{Foxtwo_state_subs}). 
\begin{align}
   \mathbf{T}(t)=
   \begin{bmatrix}
   -k_{12} & k_{21}(t) & 0 & 0\\
   k_{12}  & -k_{21}(t)-k_{23} & k_{32} & 0\\
   0 &  k_{23} & -k_{32}-k_{34} & k_{43} \\
   0 & 0 & k_{34} & -k_{43}
   \end{bmatrix}
   ;
   \mathbf{B}=
   \begin{bmatrix}
   k_{r_{1}} & 0 & 0 & 0 \\
   0 & k_{r_{2}} & 0 & 0 \\
   0 & 0 & k_{r_{3}} & 0 \\
   0 & 0 & 0 & k_{r_{4}}
   \end{bmatrix}
   ;
   \boldsymbol{\Gamma}=
   \begin{bmatrix}
   \gamma & 0 & 0 & 0 \\
   0 & \gamma & 0 & 0  \\
   0 & 0 & \gamma & 0 \\
   0 & 0 & 0 & \gamma
   \end{bmatrix}.\label{FoxFoxElem}
\end{align}
\begin{equation}\label{FoxFoxmastermatrix}
    \frac{d}{dt}
    \begin{bmatrix}
    \mathbf{P}_0 \\
    \mathbf{P}_1 \\
    \mathbf{P}_2 \\
    \vdots \\
    \Pv_{N_m} \\
    g(t)
    \end{bmatrix}
    =
    \begin{bmatrix}
    \mathbf{T-B} & \mathbf{\boldsymbol{\Gamma}} & \mathbf{0} & \hdots & 0 \\
    \mathbf{B} & \mathbf{T}-\mathbf{B}-\boldsymbol{\Gamma} & \mathbf{2\boldsymbol{\Gamma}} & \ddots & 0 \\
    \mathbf{0} & \mathbf{B} & \mathbf{T}-\mathbf{B}-2\boldsymbol{\Gamma} & \ddots & 0 \\
    \vdots & \ddots & \ddots & \ddots & 0 \\
    \mathbf{0} & \hdots & \mathbf{B} & \mathbf{T}-\mathbf{B}-N_m\boldsymbol{\Gamma} &  0 \\
    \mathbf{0} & \hdots & \hdots  &\mathbf{1}^T \mathbf{B} & 0
    \end{bmatrix}
    \begin{bmatrix}
    \mathbf{P}_0 \\
    \mathbf{P}_1 \\
    \mathbf{P}_2 \\
    \vdots \\
    \Pv_{N_m} \\
    g(t)
    \end{bmatrix},
\end{equation}
This model suggests that \textit{STL1} and \textit{CTT1} expression is activated from a low gene expression state (OFF, in Figure \ref{FoxHogFig}(B)), to higher gene expression states by modulation of the return rate to the off state, $k_{21}$.
When Hog1-p levels in the nucleus are above a certain threshold (denoted by the cyan line in Figure \ref{FoxHogFig}(A), the S2 state becomes more stable and allows more RNA to be created by transitioning into the more highly active S3 and S4 states. 

\section{Summary}
Noise in biological systems can have huge implications about how we understand biological behaviors, from the creation and destruction of single biochemical species to populations of cells, to entire organisms. To develop mathematical models that help us to understand gene expression dynamics, analyses are needed that capture the discrete stochastic behaviors of genes, RNA and protein.  One such modeling approach is the CME, in which gene expression is represented by a discrete state Markov chain.  The CME describes the time evolution of probabilities of different biochemical species according to their corresponding reactions and stoichiometries.  However, because the CME often has infinite dimension, it is not directly solvable, either computationally or analytically.  The FSP approach selects a subset of the CME dimension, where the complement set of states is combined into one or more absorbing sink states.  Because the FSP approach includes a finite number of states, the time evolution of those states can be solved by integrating a finite set of ODEs, and the sinks quantify the precise error in the FSP approximation at any finite instance in time. For systems with less than four interacting species whose population are relatively small (approximately less than $1000$), the FSP approach can be computationally efficient in finding such distributions with guaranteed accuracy. However, for large dimension systems, it remains to be seen if FSP approximations can suffice to solve the CME.

The key advantage of the CME/FSP formulation of gene regulation models is that they can be matched easily to single-cell data. Using discrete stochastic models is particularly useful when the data being considered is also discrete (i.e.\ RNA or protein molecule counts in cell) and heterogeneous across a population of cells. The likelihood of observing data for FSP models can be derived to match FSP analyses to such discrete stochastic data. Moreover, the FSP error may be used to find precise bounds on this likelihood, which helps us to understand the tradeoff between computation expense (FSP error) and model identification.
FSP computed likelihoods have been used to match experimentally measure smFISH data in bacteria, yeast, and human cells. We note that the dependence of single-cell distributions on discrete events (e.g., single-molecule reactions) leads to complex distributions that may have either multiple peaks, as in Figure \ref{FoxHogFig} or long tails as in Figures \ref{FoxHogFig} and \ref{FoxYSP8}.  For such distributions, standard assumptions about the underlying stochastic processes, such as those based upon the central limit theorem, may not be sufficient to identify models. For example, a common approach is to use the first two moments from the CME and compare them to the measured moments of the data under the assumption of Gaussian fluctuations. In such cases, and without sufficient data to invoke the central limit theorem, moments-based analyses may fail, while more detailed CME/FSP analyses, which use all of the fluctuations within the data, may be capable to identify model parameters with more precision and greater predictive accuracy. 

\section*{Acknowledgments}  BM and ZF were funded by the W.M. Keck Foundation, DTRA FRCALL 12-3-2-0002 and CSU Startup Funds. 


\clearpage
\section*{Supplemental Exercises}\label{exercises}

The interested reader is encouraged to test their understanding of the presented material through completion of the following exercises.

\begin{enumerate}
\item Use the results derived in Example \ref{exercises}.2 above to explain why systems with large number of molecules tend to have lower levels of variability.
\item Use the definition of the Poisson distribution to show that the steady state mean and variance are equivalent in the housekeeping gene transcription model.
\item Consider the following two reactions:
\begin{align*}
    \mathcal{R}1: & mRNA \xrightarrow{k_1} mRNA + Protein \\
    \mathcal{R}2: & mRNA \xrightarrow{k_2} \varnothing
\end{align*}
Starting with 1 mRNA at $t=0$, answer the following questions: 
    \begin{enumerate}
        \item Find the probability that $\mathcal{R}2$ occurs before $\mathcal{R}1$.
        \item Find the probability that $\mathcal{R}1$ occurs at least once before $\mathcal{R}2$ does.
        \item Find the probability that $\mathcal{R}1$ occurs exactly $n$ times before $\mathcal{R}2$ does. 
    \end{enumerate}
\item A transcription factor may bind to a gene in two different sites, and can therefore create three different binding configurations of the gene (see figure below). If the gene starts in state G1 at $t=0$, find the probability that the gene is in state $G3$ at time $t$. Use the transition rates given in the figure. 
\item You observe a protein that is normally absent, but has occasional spikes in population level. Each spike corresponds to the sudden level of 500 proteins on average, and lasts 1 second on average. Spikes are separated by 100 seconds on average. Using a simple model of transcription and translation:
\begin{align*}
    \varnothing & \xrightarrow{k_r} mRNA \\
    mRNA & \xrightarrow{\gamma_r} \varnothing \\
    mRNA & \xrightarrow{k_p} mRNA + Protein \\
    Protein & \xrightarrow{\gamma_p} \varnothing,
\end{align*}
find a set of parameters $\boldsymbol{\Lambda} = [k_r,\gamma_r,k_p,\gamma_p]$ that could account for these observations. Note that this parameter set is not necessarily unique. 
\item Prove that the Eqn.\ \ref{Foxlikelihood} provides a lower bound on the true likelihood of the experimental data given the true CME model.
\item Show that maximization of the likelihood of the data (Eqn.\ \ref{Foxlhood}) and the minimization of the KLD (Eqn.\ \ref{FoxKLD}) occur at the same parameter set $\mathbf{\Lambda}$. 
\item Consider a system where chemical species $x$ is created, but is not able to be degraded,
\begin{align*}
    \varnothing \xrightarrow{k} x.
\end{align*}
For rate $k=10s^{-1}$, use the FSP approach to find the probability that $x$ is greater than 100 as a function of time (i.e., find $P_{>100}(t) \equiv \sum_{x=101}^{\infty} P(x,t)$). 
\item Consider the following set of parameters for the genetic toggle switch presented in Example \ref{FoxToggle}: 
\begin{align*}
   \gamma_{\xi_1} = \gamma_{\xi_2} = \eta_{\xi_1} = \alpha_{\xi_1} = \alpha_{\xi_2} = 1; \hspace{.25cm} \eta_{\xi_2} = 2.5; \hspace{.25cm} k_{\xi_1} = 50; \hspace{.25cm} k_{\xi_2}=16; \hspace{.25cm} b_{\xi_1} = b_{\xi_2} = 0, 
\end{align*}
and the initial condition $\xv(0) = \left[ 0 \hspace{.2cm} 0 \right]$. Consider the system to be ON if $\xi_1>15$ and OFF if $\xi_2>5$, and undetermined otherwise. First, simulate many trajectories of this system using the Stochastic Simulation Algorithm (see Chapter 7 of \cite{Munsky:2017MIT}) and do the following:
\begin{enumerate}
    \item Plot one trajectory $\xi_1$ and $\xi_2$ vs. time.
    \item For each run, determine the time at which the switch first turns OFF and the time at which the switch first turns ON. 
    \item What are the median times at which these first switches occur? 
    \item What portion of runs first turns OFF before turning ON? 
\end{enumerate}
\item Use the Finite State Projection approach to analyze the system from problem \ref{exercises}.9. For your projection, define constraints such that $b_1 = 260$, $b_2 = 40$, and $b_3=100$. Do the following:
\begin{enumerate}
   \item Plot the marginal probability distributions for $\xi_1$ and $\xi_2$ and then as a contour plot. 
   \item Change the projection to include all configurations such that $\xi_1 \leq 15$  and $\xi_2 \leq 40$ (i.e.\ $b_2=15$, $b_3=40$). Use this projection to find the times at which $50\%$ and $99\%$ of trajectories will turn ON.
   \item Change the projection to include all configurations such that $\xi_1 \leq 100$ and $\xi_2 \leq 5$. Use this projection to find the times at which $50\%$ and $99\%$ of the trajectories will turn OFF. 
   \item Use another projection to compute the probability that a cell will turn OFF before it will turn ON. 
   \item Compare your results to those found with the SSA.
\end{enumerate}
\item Use the parameters in the caption of Fig.\ \ref{Foxtoggle_demo} to recreate the marginal distributions and joint distributions (contour plot) shown in Fig.\ \ref{Foxtoggle_demo}B,C.
\end{enumerate}

\end{document}